\documentclass[final,english]{elsarticle}
\usepackage[T1]{fontenc}
\usepackage[latin9]{inputenc}
\usepackage{varioref}
\usepackage{units}
\usepackage{amsmath}
\usepackage{amssymb}
\usepackage{graphicx}
\usepackage{esint}
\usepackage{caption}
\usepackage{subcaption}

\pdfpageheight\paperheight
\pdfpagewidth\paperwidth

\journal{Elsevier}
\usepackage{upgreek}
\usepackage[bookmarks,bookmarksopen,bookmarksdepth=6]{hyperref}
\usepackage{cases}
\usepackage{url}
\usepackage{lineno}

\usepackage{a4wide}

\usepackage{newtxtext,newtxmath,amsmath}
\makeatother

\usepackage{babel}
\begin{document}

\title{Is the gain-voltage dependence of SiPMs linear?}
\author[]{M.~Antonello}
\author[]{L.~Brinkmann}
\author[]{E.~Garutti}
\author[]{R.~Klanner\corref{cor1}}
\author[]{J.~Schwandt}

\address{Institute for Experimental Physics, University of Hamburg,
 \\Luruper Chaussee 149, 22761, Hamburg, Germany.}

\cortext[cor1]{Corresponding author. Email address: Robert.Klanner@desy.de}


\begin{abstract}

The gain-voltage dependence for SiPMs from Ref.\,\cite{Chmill:2017} (V.\,Chmill et al., Study of the breakdown voltage of SiPMs) is reanalyzed and a non-linearity at the sub-percent level is observed.
Simulations show that the non-linearity can be explained by the increase of the depletion depth of the avalanche region with over-voltage. 
 A consequence of the non-linearity is that the voltage at which the discharge stops is systematically underestimated if a linear extrapolation is used. 
However, the shift is too small to explain the difference between the break-down voltage from the current-voltage dependence and the one obtained from the extrapolation of the gain-voltage dependence to gain one. 
The results are confirmed by measurement of a different MPPC produced by Hamamatsu which has a break-down voltage which is about a factor two higher.

\end{abstract}


\maketitle
 \pagenumbering{arabic}

\section{Introduction}
 \label{sect:Introduction}

Silicon Photo-Multipliers (SiPMs) are arrays of avalanche photo-diodes, called pixels in this paper, operated above the break-down voltage.
 Typical pixel pitches are between 10 and 50\,$\upmu$m with sensitive SiPM areas of 1 to 6\,mm$^2$.
 Thanks to their excellent performance, single photon detection with a gain, $G$, of order $10^6$, sub-nanosecond timing resolution, high photon-detection efficiency, operating voltages below 100\,V, insensitivity to magnetic fields and moderate cost, they have replaced vacuum photo-multipliers in many applications. 

It has been observed that, as long as the probability that more than one photon converts in the active region of a pixel is low, $G$ increases to a good approximation linearly with the bias voltage $U_b$ above the break-down voltage $U_\mathit{bd}$.
Extrapolating this linear dependence to $G=1$, allows to estimate $U_\mathit{off}$, the voltage at which the discharge stops.
They are not necessarily the same, however it is observed  that $U_\mathit{off} \approx U_\mathit{bd}$.
This paper discusses how well the linearity is satisfied and how valid the linear extrapolation is for determining $U_\mathit{off}$.

 Evidence for a non-linear gain-voltage dependence is reported in Refs.\cite{Acerbi:2015, Acerbi:2019}, and explained by \emph{"This is due to the progressive depletion of the epi-layer beneath the p-n junction, leading to a diode capacitance reduction with increasing bias voltage, thus a non linear gain dependence."}
The study presented here confirms the observation and the explanation.

The structure of the paper is as follows:
The next section presents 1D electric field calculations as a function of $U_b$ for a $p$-$n$\,pixel using a doping profile which is realistic for the center of a pixel. 
As expected, the depletion depth increases with $U_b$, which results in a decrease of the pixel capacitance, $C_\mathit{pix}$.
From the formula (which ignores the effect of an additional capacitance parallel to the quenching resistor)
  \begin{equation}\label{equ:G_Ub}
G(U_b) = \max\left(1,\frac{  C_\mathit{pix}(U_b) \cdot U_b - C_\mathit{pix}(U_\mathit{bd}) \cdot U_\mathit{bd} } {q_0} \right) 
 \end{equation}
with the elementary charge $q_0$, it can be seen that a voltage-dependent $C_\mathit{pix}$ results in a non-linear dependence of $G(U_b)$.
If $C_\mathit{pix}$ does not change for $U_b \geq U_\mathit{bd}$, Eq.\,\ref{equ:G_Ub} simplifies to $G(U_b) = \max( 1,C_\mathit{pix} \cdot (U_b - U_\mathit{bd}) / q_0)$, which has been used so far in the literature. 
In the following section data for KETEK SiPMs with pixel pitches of 15 and 25\,$\upmu$m from Ref.\,\cite{Chmill:2017} show that a non-linearity similar to the one predicted by the simulation is observed. 
Data from a MPPC HPK\,13360\,-1325 fabricated by Hamamatsu confirm these results.
In addition, the differences of $U_\mathit{off}$ using a linear and a quadratic fit to $G(U_b)$ for the extrapolation to $G = 1$ are presented.
The last section summarizes the main results. 

 \section{Simulation of the voltage dependence of the gain}
  \label{sect:Simulation}

To estimate the electric field in a SiPM pixel a 1D doping profile  has been simulated using SYNOPSYS TCAD\,(\cite{Synopsys}). 
The process simulation, which uses confidential information provided by KETEK\,\cite{KETEK}, includes phosphorous and boron implantations as well as annealing steps.
The doping profile for the donors is denoted $N_D(x)$, for the acceptors $N_A(x)$, and the difference  is $N_d(x) = N_D(x) - N_A(x)$; $x = 0$ is the position of the SiPM entrance window. 


Next, the calculation of the dependence of the pixel capacitance on bias voltage, $C_\mathit{pix} (U_b)$,  is presented.
The position of the junction, $x_j$, is obtained from $N_d(x_j) = 0$, where the value $x_j = 0.55\,\upmu$m is found.
In the further calculations, the width of the $P$-doped depletion region, $d_n$, is used as free parameter.
For a given $d_n$, the boundary of the $B$-doped depletion region, $x_p$, is then calculated using the neutrality condition, $\int_{x_p} ^{x_j+d_n} N_d(x)\,dx = 0$.
From the boundaries of the depletion region, the absolute value of the electric field, $E(x;d_n) = (q_0/\varepsilon_\mathrm{Si})\cdot \int_{x_p} ^{x} N_d(x')\,dx'$, is obtained.
Finally, the bias voltage, $U_b(d_n)$ is calculated by integrating the electric field over the depletion region. 
The break-down voltage, $U_\mathit{bd}$, is the value of $U_b(d_n)$ for which the ionization integral is 1\,\cite{Sze:1981}.
For the ionization coefficients, the data of Ref.\,\cite{Overstraeten:1970} at 300\,K have been used.
The value obtained for $U_\mathit{bd}$ is 27.85\,V at a depletion depth $d_\mathit{bd} = 0.879\,\upmu $m. 

\begin{figure}[!ht]
  \centering 
   \begin{subfigure}[a]{0.5\textwidth}
   \includegraphics[width=\textwidth]{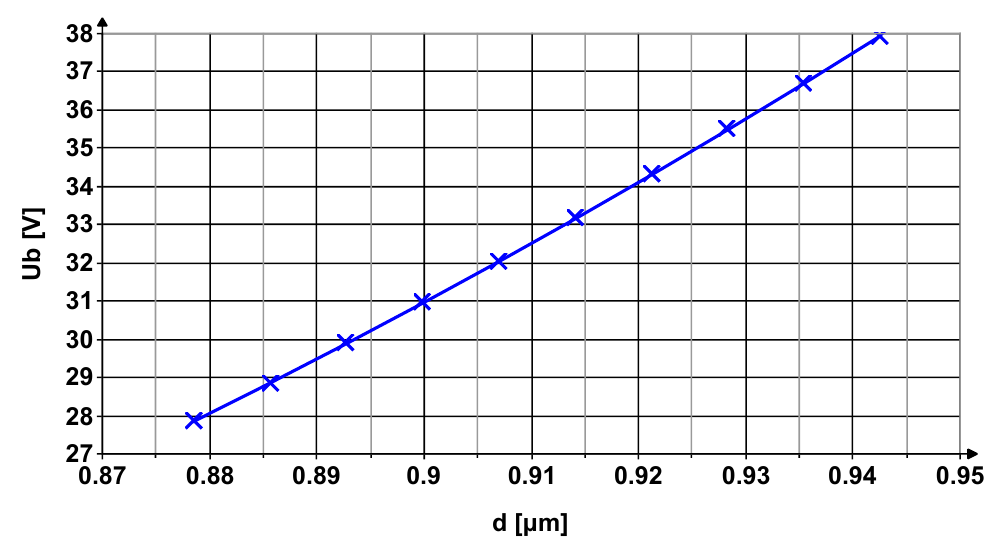}
  \caption{ }
   \label{fig:depl}
 \end{subfigure}%
   ~
 \begin{subfigure}[a]{0.5\textwidth}
   \includegraphics[width=\textwidth]{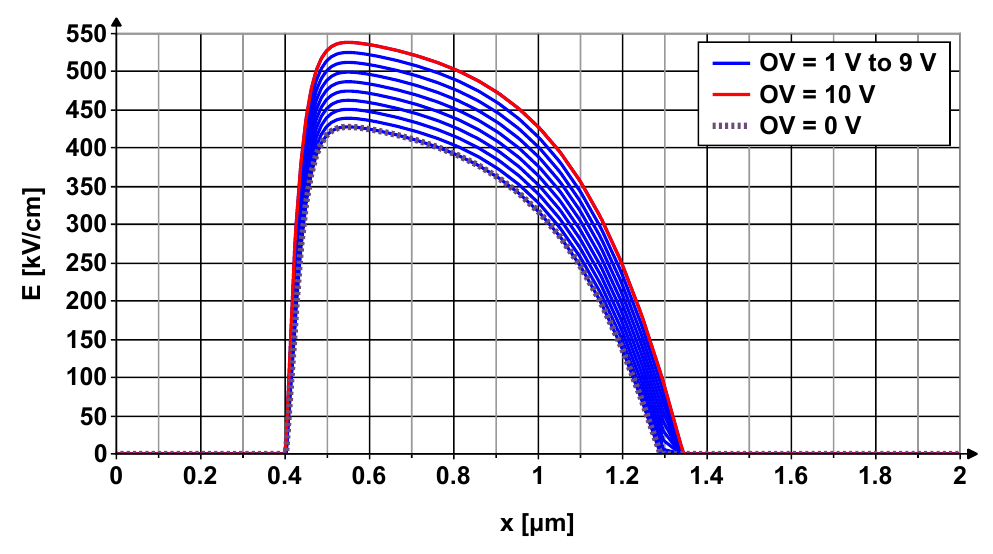}
  \caption{ }
   \label{fig:Efield}
 \end{subfigure}%
  \caption{ 
	  (a) Simulated dependence of the bias voltage, $U_b$, on the depletion depth $d$.
		(b) Distribution of the absolute value of the electric field, $E(x)$, for 10 over-voltages between 0 and 10\,V.}
  \label{fig:Edepl}
 \end{figure}

Fig.\ref{fig:depl} shows the relation between the bias voltage, $U_b$, and the total depletion depth, $d$, and Fig.\ref{fig:Efield} the position dependence of the electric field for over-voltages, $\mathit{OV} = U_b - U_\mathit{bd}$, between 0 and 10\,V.
It can be seen that the depletion depth increases with \emph{OV}.
As a consequence, the pixel capacitance, $C_\mathit{pix}$, decreases with \emph{OV}, which, according to Eq.\,\ref{equ:G_Ub}, results in a non-linear $G(\mathit{OV})$ dependence.  

\begin{figure}[!ht]
  \centering 
   \begin{subfigure}[a]{0.5\textwidth}
   \includegraphics[width=\textwidth]{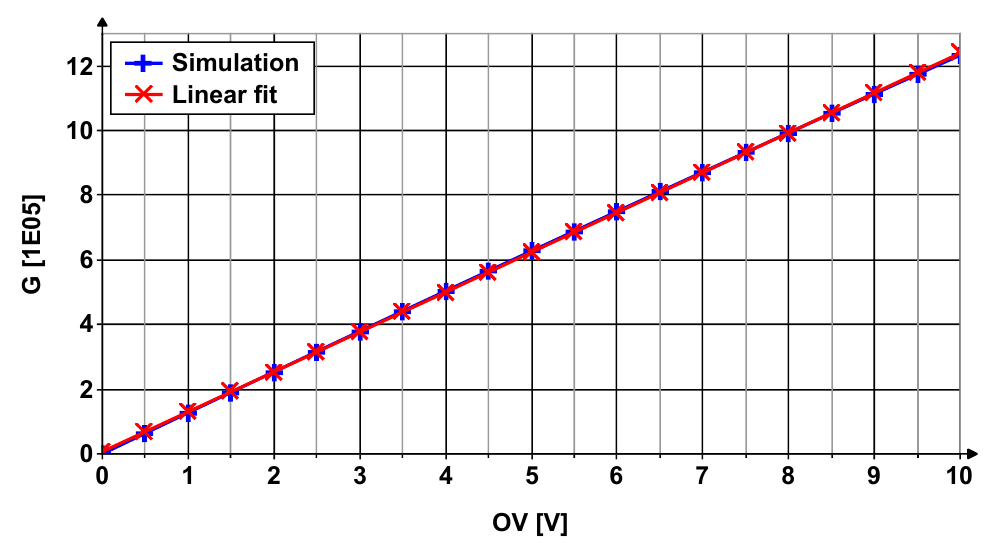}
  \caption{ }
   \label{fig:Gsim}
 \end{subfigure}%
   ~
 \begin{subfigure}[a]{0.5\textwidth}
   \includegraphics[width=\textwidth]{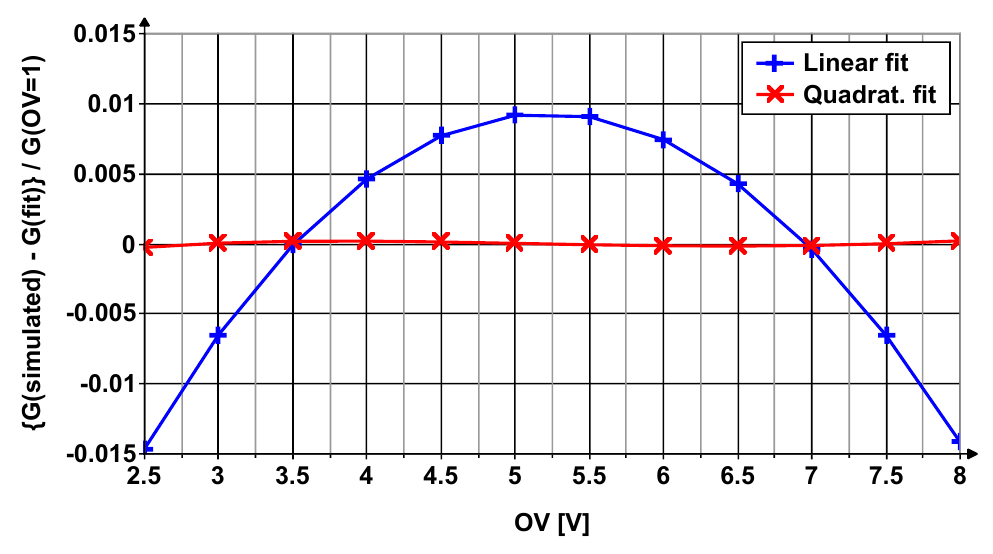}
  \caption{ }
   \label{fig:dGsim}
 \end{subfigure}%
  \caption{ (a) Simulated gain as a function of over-voltage for a pixel pitch of 15\,$\upmu$m. 
	   (b) Difference between the data shown in (a) to a linear and a quadratic fit divided by the gain at $\mathit{OV} = 1$\,V.}
  \label{fig:Gainsim}
 \end{figure}

For a depletion depth, $d$, $C_\mathit{pix} \approx \varepsilon_\mathrm{Si} \cdot a^2 / d$, and the gain $G(\mathit{OV})$ can be calculated  using Eq.\,\ref{equ:G_Ub}.
The result for a square pixel of size $a=15\,\upmu$m is shown in Fig.\,\ref{fig:Gsim} together with a linear fit in the \emph{OV} interval between 0 and 10\,V. 
The fit appears to describe the data, however, there are small systematic deviations, which are hardly visible in the figure.
The data discussed in section\,\ref{sect:Measurement} consist of 12 gain measurements for over-voltages between about 2.5 and 8.0\,V. 
Fig.\,\ref{fig:dGsim} shows for this \emph{OV}\,interval the differences between the simulations and the results of a linear and a quadratic fit divided by the gain for $\mathit{OV} = 1$\,V. 
For the linear fit an approximately parabolic deviation with values between $ - 0.015$ and $ + 0.009$ is observed, whereas the parabolic fit shows deviations below $5 \times 10^{-4}$.
It is noted that a deviation divided by $G(\mathit{OV} = 1$\,V) of 0.01 at $\mathit{OV} = 5$\,V corresponds to a relative gain difference of 0.2\,\%.
Thus, there are small, but significant deviations from a straight-line fit.
More relevant is that the straight line extrapolated to $G=1$ differs from $U_\mathit{bd}$. 
For the data shown in Fig.\,\ref{fig:dGsim} the difference is $ -88$\,mV which significantly exceeds the uncertainty of the extrapolation.
For the parabolic fit the difference is $-6.4$\,mV and thus much smaller.
It is noted that the difference for the linear fit also depends  on the \emph{OV}\,interval fitted and, for a given \emph{OV}\,interval, on the distribution and the number of data points. 
For the \emph{OV}\,interval between 2.5 and 8\,V, $U_\mathit{off}$ changes by about 5\,mV when the number of data points is changed from 5 to 100.
Assuming a quadratic dependence of $G(\mathit{OV})$ and equal uncertainties of the individual $G$\,measurements, the difference of the linear extrapolation to $G = 1$, can be estimated by 
\begin{equation}
 \label{equ:dUoff}
  \Delta U_\mathit{off} =  \langle \mathit{OV} \rangle  - \langle G \rangle /b,
\end{equation}
where $b$ is the slope of the linear fit to the $G(U_b)$\,data, and $\langle \rangle$ are the means for the data points. 

To summarize the results of the simulation presented in this section: 
The depletion depth of the avalanche photo-diodes of  SiPMs increases with over-voltage, which results in a non-linear gain--over-voltage dependence. 
The non-linearity is small, but, using a linear extrapolation results in a negative shift of the value determined for $U_\mathit{off}$, the voltage at which the discharge stops.  
For the simulation presented it can be shown that a quadratic over-voltage dependence provides an adequate parametrization and its extrapolation gives $U_\mathit{off} = U_\mathit{bd}$ within less than 10\,mV.
  
\section{Measured voltage dependence of the gain}
 \label{sect:Measurement}
 
The data presented in Ref.\,\cite{Chmill:2017} have been used for the study of the non-linearity of the gain as a function of over-voltage.
One reason for this choice has been the observation of the pixel pitch dependence of the difference between the break-down voltage determined from current measurements $(I-V)$ and the voltage at which the discharge stops obtained from a linear extrapolation, which is so far not understood.
One motivation for the present study has been to find out if a non-linearity can explain this observation. 

\begin{table}[h] 
    \caption{Properties of the SiPMs and measurement conditions:
		$\tau $ the pulse decay time, $U_\mathit{bd}$ the break-down voltage from $I-V$\,measurements, $T$ the temperature at which the data were taken, $t_\mathit{gate}$ the width of the QDC gate, and $\lambda_\gamma$ the wavelength of the light. }
    \label{tab:SiPMs}
		  \centering 
 \begin{tabular}{c|c|c|c|c|c|c}
     \emph{pitch} [$\upmu$m] & $N_\mathit{pix}$  & $\tau$ [ns] & $U_\mathit{bd}$ [V]  &  $T$ [$^\circ$C] & $ t_\mathit{gate}$ [ns] & $\lambda_\gamma$ [nm] \\
   \hline
     15 & 4382 & 13.5  & 27.5  & 20 & 100 & 470 \\
     25 & 2304 & 35     & 27.9  & 20 & 200 & 470 \\		
   \hline
 \end{tabular}
  \end{table}

Here only a short summary of the SiPMs and the measurement setup is given.
More details can be found in Ref.\,\cite{Chmill:2017}.
The SiPMs were produced by KETEK\,(\cite{KETEK}); Table\,\ref{tab:SiPMs} presents some of their properties.
The charge spectra were recorded using a CAEN QDC\,(\cite{CAEN}) for bias voltages $U_b = $\,29.5, 30.5 .. 35.0\,V for the SiPM with a pitch of $15\,\upmu$m and $U_b = $\,29.0, 29.5 .. 35.0\,V  for the $25\,\upmu$m  SiPM.
The gain values were obtained by fitting the charge spectra using PeakOtron\,(\cite{Rolph:2023}).

\begin{figure}[!ht]
  \centering 
   \begin{subfigure}[a]{0.5\textwidth}
   \includegraphics[width=\textwidth]{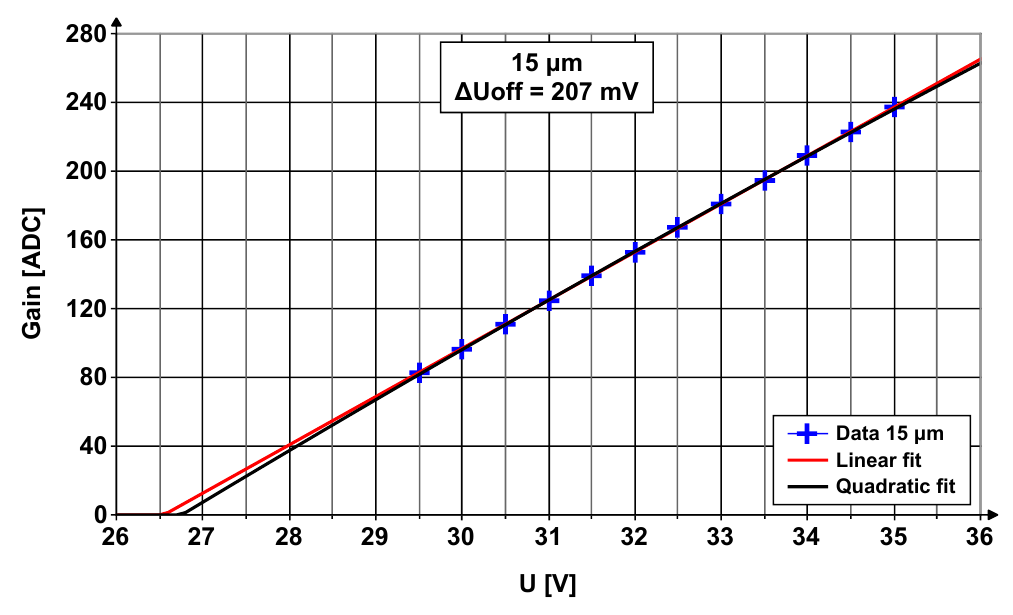}
  \caption{ }
   \label{fig:Gain15mum}
 \end{subfigure}%
   ~
 \begin{subfigure}[a]{0.5\textwidth}
   \includegraphics[width=\textwidth]{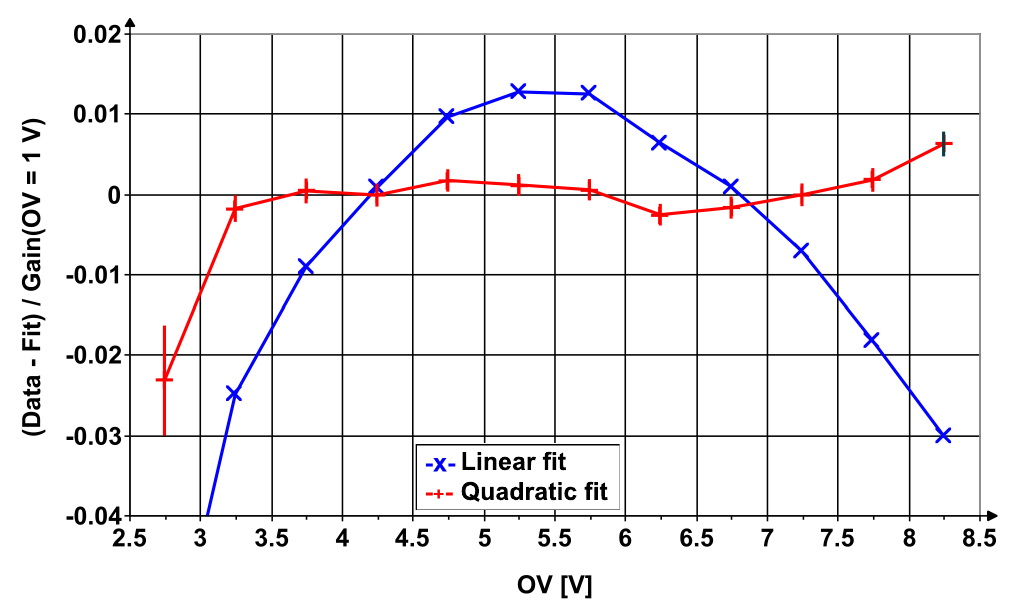}
  \caption{ }
   \label{fig:DiffGain15}
 \end{subfigure}%
  \caption{ 
	  (a) Measured gain as a function of the bias voltage for the KETEK SiPM with the $15\,\upmu$m pitch, and the results of a linear and a quadratic fit in the $U_b$-range 30.0\,V to 34.5\,V. 
		(b) Difference of the measured gain and the results of the linear and quadratic fits as a function of the over-voltage, the difference of $U_b$ and $U_\mathit{off}$ from the quadratic fit, for the data shown in (a).
	The statistical uncertainties of the gain measurements are only shown for the quadratic fit. It is the same for the linear fit. 	
		Note that the lowest \emph{OV}\,point for the linear fit is off scale.
}
  \label{fig:G15}
 \end{figure}

Fig.\,\ref{fig:Gain15mum} shows for the $15\,\upmu$m SiPM the gain-bias voltage measurements together with a linear and a quadratic fit in the $U_b$-range 30.0\,V to 34.5\,V. 
The fitted curves, extrapolated to $G=1$, give $U_\mathit{off}$. 
The values are $26.5500 \pm 0.0018$\,V and $26.7565 \pm 0.0067$\,V, respectively. 
Thus, the linear fit results in an $U_\mathit{off}$\,value which is lower by $207 \pm 7$\,mV. 
Only the statistical uncertainties are given.
The values of $U_\mathit{off}$ are dominated by systematic effects, however, it is expected that most systematic effects cancel in the $U_\mathit{off}$ difference.
It is noted that for the linear fit $\chi ^2/\mathrm{NDF} = 906/8$, and for the quadratic fit 13.7/7.
Fig.\,\ref{fig:DiffGain15} shows the differences of the measured gain and the linear and quadratic fits. 
The curve for the linear fit is approximately a parabola, similar to the expectation from the simulation shown in Fig.\,\ref{fig:dGsim}.
The quadratic fit reduces the mean deviation, data minus fitted values, by nearly an order of magnitude. 

\begin{figure}[!ht]
  \centering 
   \begin{subfigure}[a]{0.5\textwidth}
   \includegraphics[width=\textwidth]{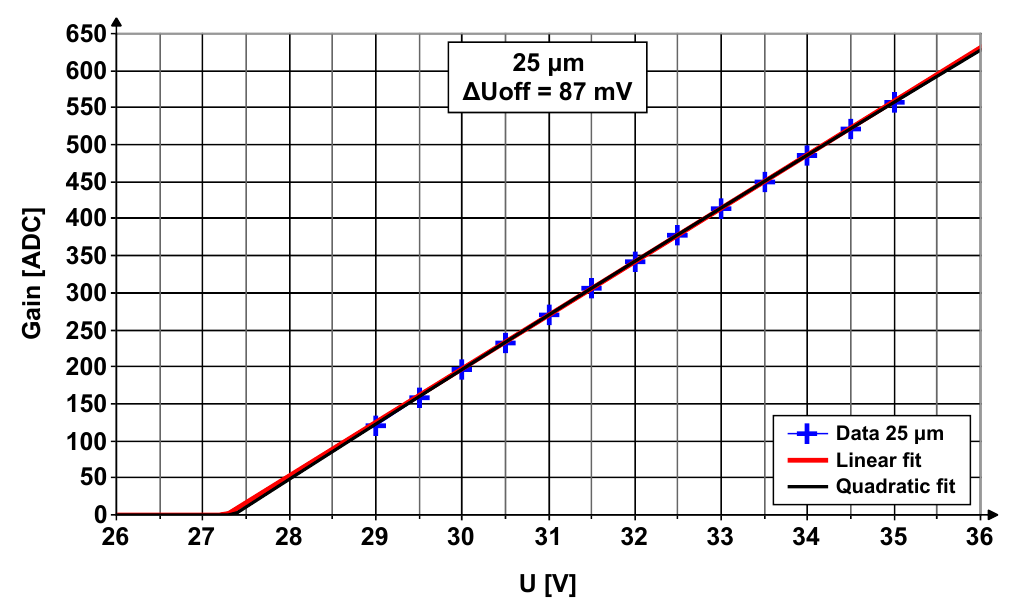}
  \caption{ }
   \label{fig:Gain25mum}
 \end{subfigure}%
   ~
 \begin{subfigure}[a]{0.5\textwidth}
   \includegraphics[width=\textwidth]{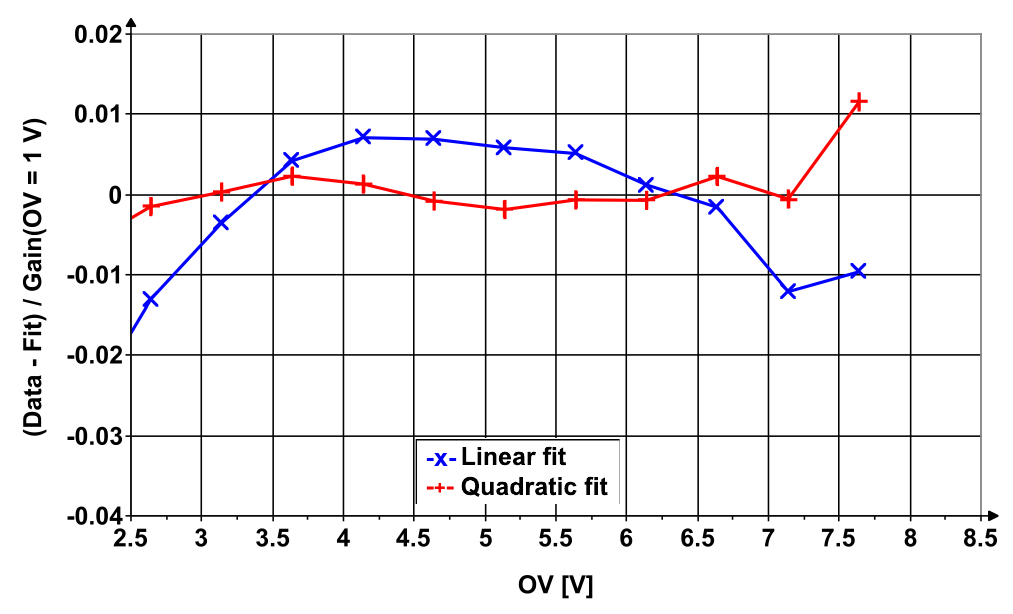}
  \caption{ }
   \label{fig:DiffGain25}
 \end{subfigure}%
  \caption{ 
	  (a) Measured gain as a function of the bias voltage for theKETEK SiPM with the $25\,\upmu  $m pitch, and the results of a linear and a quadratic fit in the $U_b$-range 30.0\,V to 34.5\,V.
		(b) Difference of the measured gain and the results of the linear and quadratic fits as a function of the over-voltage, the difference of $U_b$ and $U_\mathit{off}$ from the quadratic fit, for the data shown in (a).
		The statistical uncertainties of the gain measurements are only shown for the quadratic fit. They are the same for the linear fit. 	}
  \label{fig:G25}
 \end{figure}

Fig.\,\ref{fig:G25} shows the results corresponding to Fig.\,\ref{fig:G15} for the $25\,\upmu$m SiPM.
The values for $U_\mathit{off}$ for the linear and the quadratic fits are $27.2730 \pm 0.0013$\,V and $27.3601 \pm 0.0040$\,V, respectively, and the difference is $87 \pm 4$\,mV.
Again, only the statistical uncertainties are given.
For the linear fit $\chi ^2/\mathrm{NDF} = 507/8$, and for the quadratic fit 19.7/7.
The quadratic fit reduces the mean deviation by about a factor five.


It is concluded that for both $15\,\upmu$m and $25\,\upmu$m SiPM the $G(\mathit{OV})$ dependence is not linear. 
The difference of $U_\mathit{off}$ obtained by extrapolating the linear and the quadratic $G(U_b)$ fits to $G = 1$ is $-207\pm22$\,mV for the $15\,\upmu$m SiPM and $-87\pm12$\,mV for the $25\,\upmu$m SiPM.
In Ref.\,\cite{Chmill:2017} differences between the linear $G(U_b)$ extrapolation  and $U_\mathit{bd}$ from $I-V$ measurements for the $15\,\upmu$m SiPM of $-900\pm100$\,mV, and for the $25\,\upmu$m SiPM of $-550\pm50$\,mV are reported.
Thus, it can be concluded that only part of the observed differences can be explained by the systematic shifts due to the linear extrapolation of $G(U_b)$.

In order to check if other SiPMs also have a non-linear gain-voltage dependence, data have been taken with the MPPC HPK13360\,-1325	which has 2668 pixels with a pitch of 25\,$\upmu$m\,\cite{HPK13360}.
The charge spectra were recorded using a CAEN QDC for bias voltages $U_b = $\,53.0, 53.5 .. 60.0\,V with a gate width of 104\,ns at a temperature of about 23.5\,$^\circ$C, and the gain values were obtained by fitting the charge spectra using PeakOtron.

\begin{figure}[!ht]
  \centering 
   \begin{subfigure}[a]{0.5\textwidth}
   \includegraphics[width=\textwidth]{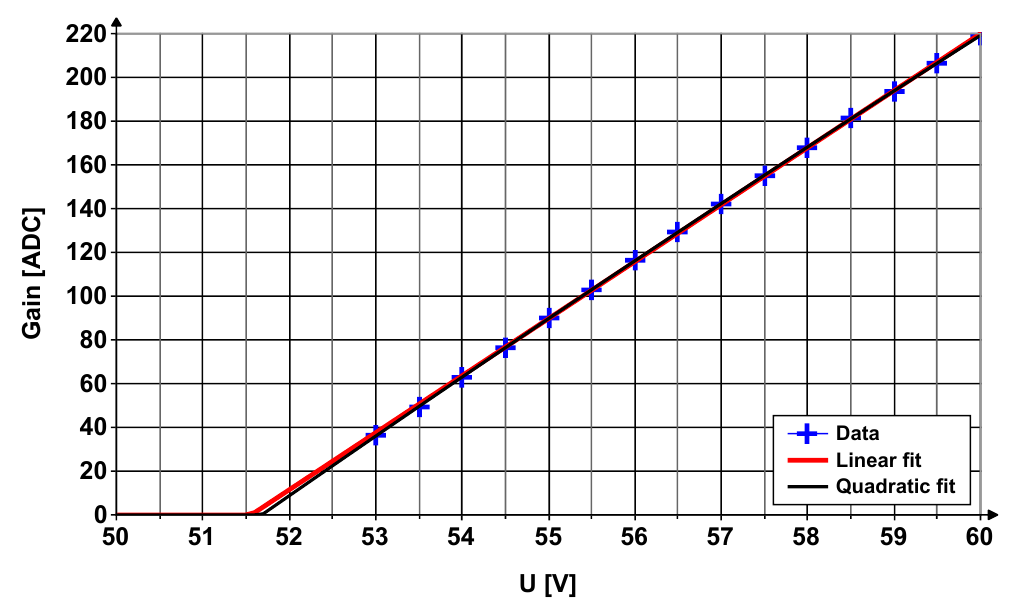}
  \caption{ }
   \label{fig:GainHPK}
 \end{subfigure}%
   ~
 \begin{subfigure}[a]{0.5\textwidth}
   \includegraphics[width=\textwidth]{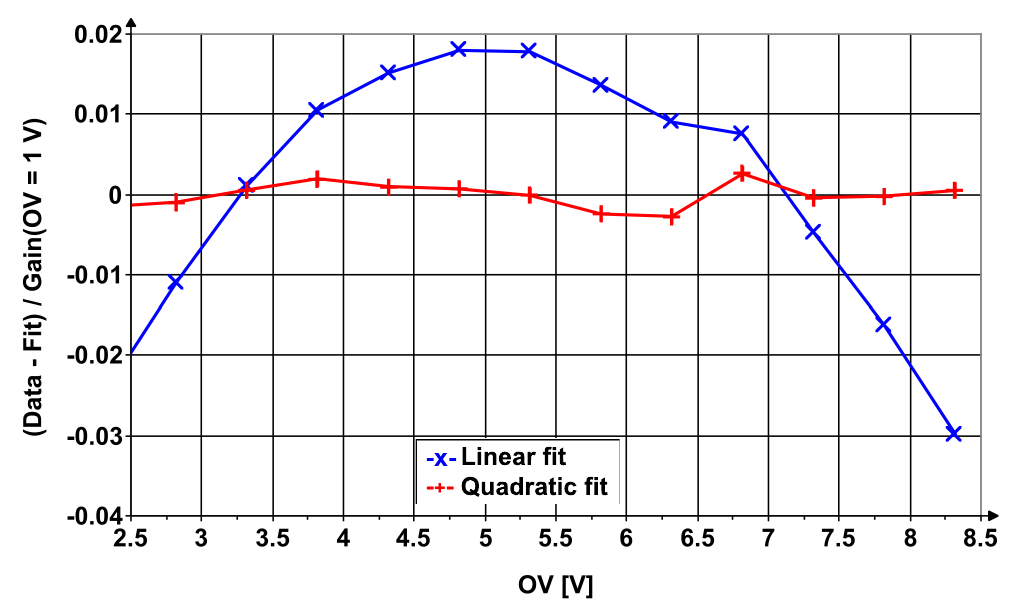}
  \caption{ }
   \label{fig:DiffGainHPK}
 \end{subfigure}%
  \caption{ 
	  (a) Measured gain as a function of the bias voltage for the MPPC HPK13360\,-1325 and the results of a linear and a quadratic fit. 
		(b) Difference of the measured gain and the results of the linear and quadratic fits as a function of the over-voltage, the difference of $U_b$ and $U_\mathit{off}$ from the quadratic fit, for the data shown in (a).
 	The statistical uncertainties of the gain measurements are only shown for the quadratic fit. They are the same for the linear fit. 		
	}
  \label{fig:HPK}
 \end{figure} 

Fig.\,\ref{fig:GainHPK} shows the gain as a function of the bias voltage and the results of a linear and a quadratic fit. 
The values of $U_\mathit{off}$, obtained by extrapolating the fit curves to $G = 1$, are $51.5606 \pm 0.0007$\,V and $51.6828 \pm 0.0016$\,V respectively, and the difference is $122 \pm 2$\,mV.
The corresponding $\chi ^2$/\emph{NDF} values are 5997/13 and  51/12.
Only the statistical uncertainties are given.
It can be concluded that also this SiPM shows a non-linear gain-voltage dependence and that the values of $U_\mathit{off}$ differ if a linear or quadratic fit is used.
Fig.\,\ref{fig:DiffGainHPK} shows the differences of the measured gain and the linear and the quadratic fits. 
The curve for the linear fit is approximately a parabola. 
The quadratic fit reduces the mean deviation by about an order of magnitude. 

It can be concluded that the MPPC HPK13360\,-1325, with a break-down voltage of about 52\,V and the KETEK SiPMs with a break-down voltage of about 28\,V, show similar non-linearities.
In both cases a quadratic dependence of the gain on over-voltage provides a fair description of the data. 

  \section{Conclusions}
   \label{sect:Summary}

For two SiPMs produced by KETEK with pixel pitches of $15\, \upmu$m  and $25\, \upmu$m it is observed that the dependence of the gain on over-voltage is non-linear.
The non-linearity, which depends on the over-voltage values of the measurements, is small, typically well below 1\,\%.
However, when $U_\mathit{off}$, the voltage at which the discharge stops is determined by extrapolating a linear or a quadratic fit to the $G(\mathit{OV})$ data, significantly different $U_\mathit{off}$ values are obtained.
The difference of $U_\mathit{off}$ between the linear and the quadratic fits only explains part of the difference between the break-down voltage obtained from $I-V$ measurements and $U_\mathit{off}$ reported in Ref.\,\cite{Chmill:2017}. 
A 1D simulation with a realistic doping profile shows that the non-linearity can be explained by the increase of the depletion depth of the avalanche region with over-voltage. 
Data from the MPPC HPK13360\,-1325, which has a break-down voltage which is about a factor of two higher than the one of the KETEK SiPMs, show a similar non-linearity. 

 \section*{Acknowledgements}
 \label{sect:Acknowledgement}

 We thank F.\,Wiest and his colleagues from KETEK for providing the SiPMs.
Part of the work was supported by the Deutsche Forschungsgemeinschaft (DFG, German Research Foundation) under Germany's Excellence Strategy -- EXC 2121 "Quantum Universe" -- 390833306.


\end{document}